\documentclass[reprint,amsmath,amssymb,aps,]{revtex4-1}
\usepackage{graphicx}
\usepackage{sidecap}
\usepackage{array}
\usepackage{subfigure}
\usepackage{color}
\usepackage{amsmath}

\begin{document}

\title{Measurement of elliptic flow of light nuclei at $\sqrt{s_{NN}}$
 = 200, 62.4, 39, 27, 19.6, 11.5, and 7.7 GeV at RHIC}

\author{
L.~Adamczyk$^{1}$,
J.~K.~Adkins$^{20}$,
G.~Agakishiev$^{18}$,
M.~M.~Aggarwal$^{31}$,
Z.~Ahammed$^{49}$,
I.~Alekseev$^{16}$,
A.~Aparin$^{18}$,
D.~Arkhipkin$^{3}$,
E.~C.~Aschenauer$^{3}$,
A.~Attri$^{31}$
% ...
} 

% ...
\address{$^{1}$AGH University of Science and Technology, FPACS, Cracow 30-059, Poland}
% ...
\address{$^{3}$Brookhaven National Laboratory, Upton, New York 11973}
% ...
\address{$^{16}$Alikhanov Institute for Theoretical and Experimental Physics, Moscow 117218, Russia}
% ...
\address{$^{18}$Joint Institute for Nuclear Research, Dubna, 141 980, Russia}
% ...
\address{$^{20}$University of Kentucky, Lexington, Kentucky, 40506-0055}
% ...
\address{$^{31}$Panjab University, Chandigarh 160014, India}
% ...
\address{$^{49}$Variable Energy Cyclotron Centre, Kolkata 700064, India}
% ...

\collaboration{STAR Collaboration}\noaffiliation

 \begin{abstract}
 We present measurements of 2$^{nd}$ order azimuthal  anisotropy 
 ($v_{2}$) at mid-rapidity $(|y|<1.0)$ for light nuclei d, t, $^{3}$He
 (for $\sqrt{s_{NN}}$ = 200, 62.4, 39, 27, 19.6, 11.5, and 7.7 GeV)
 and anti-nuclei $\overline{\rm d}$ ($\sqrt{s_{NN}}$ =  200, 62.4, 39,
 27, and 19.6 GeV) and $^{3}\overline{\rm He}$ ($\sqrt{s_{NN}}$ = 200
 GeV) in the STAR (Solenoidal Tracker at RHIC) experiment. The $v_{2}$
 for these light nuclei produced in heavy-ion collisions is compared
 with those for p and $\overline{\rm p}$. We observe mass ordering
 in nuclei $v_{2}(p_{T})$ at low transverse momenta ($p_{T}<2.0$
 GeV/$c$). We also find a centrality dependence of $v_{2}$ for
 d and $\overline{\rm d}$. The magnitude of $v_{2}$ for t and 
 $^{3}$He agree within statistical errors. Light-nuclei $v_{2}$ 
 are compared with predictions from a blast wave model. 
 Atomic mass number ($A$) scaling of light-nuclei $v_{2}(p_{T})$ seems to
 hold for $p_{T}/A < 1.5$ GeV/$c$. Results on light-nuclei $v_{2}$
 from a transport-plus-coalescence model are consistent with
 the experimental measurements. 
 \end{abstract}

 \maketitle

 \section{{Introduction}}

 One of the main goals of high energy heavy-ion
 collision experiments is to study phase structures in the QCD
 phase diagram~\cite{QCD_theo, whitepapers}. With this purpose the
 Relativistic Heavy Ion Collider (RHIC) has finished the first phase
 of the  Beam Energy Scan (BES) program~\cite{bes_res1,
 v2_BES_prc,v2_BES_prl,v2_PID_cent, BES_PHENIX1,
 BES_PHENIX2,v2_200_QM}. It
 has been found that the identified hadron $v_{2}$ shows
 number-of-constituent-quark (NCQ) scaling at high $p_{T}$ at the
 higher beam energies. This scaling behavior is an expected signature
 of partonic collectivity via quark coalescence in the strongly
 interacting medium of quarks and gluons formed in heavy-ion
 collisions~\cite{part_coll2,part_coll3,part_coll4,part_coll1,part_Lacey}. 
 Such a  scaling behavior also suggests partonic coalescence to be a
 mechanism for hadron formation~\cite{part_coll2,part_coll3,hadr_form}.
 In a relativistic heavy-ion collision, light (anti-)nuclei can be formed
 by coalescence of produced (anti-)nucleons or from  transported
 nucleons~\cite{nucl_prod1, nucl_prod2, nucl_prod3}. The 
 binding energies of light nuclei are very small ($\sim$ few MeV),
 making it likely that surviving light nuclei are formed at a
 later stage of the evolution. This phenomenon is called
 final-state coalescence~\cite{nucl_prod1, nucl_prod4}. The
 coalescence probability of two nucleons is related to the local
 nucleon density~\cite{nucl_prod1,nucl_prod2, nucl_prod5}. Since the
 coalescence mechanism works best at the low density limit, low relative
 production of nucleons in heavy-ion collisions offers an ideal
 situation to study light-nuclei production via coalescence. 
 Measurements of azimuthal anisotropy of light nuclei offers a
 tool to understand the light-nuclei production mechanism and
 freeze-out properties at a later stage of the evolution. Unlike the
 case of quark coalescence, in a nucleon coalescence, the momentum
 space distributions of both the constituents and the products are
 measurable in heavy-ion collision experiments. 

 Prior measurements of elliptic flow $(v_{2})$ of light nuclei 
 have been carried out at the top RHIC energy ($\sqrt{s_{NN}} =$ 200
 GeV) by the PHENIX~\cite{Phenix_nucl} and the
 STAR~\cite{zhangbu_nucl,cjena_nucl} experiments. The PHENIX 
 Collaboration has measured the $v_{2}$ of deuterons (d) and
 anti-deuterons ($\overline{\rm d}$) at intermediate transverse
 momenta (1.1 $ < p_{T}< $ 4.5 GeV/$c$). How the $v_{2}$ of
 these light nuclei scale with those of (anti-)protons also has been
 reported~\cite{Phenix_nucl}. The STAR collaboration has measured the
 $v_{2}$ of d, $\overline{\rm d}$, $^{3}$He, and $^{3}\overline{\rm He}$ 
 in Au+Au collisions at $\sqrt{s_{NN}} =$ 200 GeV in the years
 2004~\cite{zhangbu_nucl} and 2007~\cite{cjena_nucl}. Negative 
 $v_{2}$ for $\overline{\rm d}$ at low $p_{T}$ has also been
 reported~\cite{zhangbu_nucl}. 
 
 In this work we expand upon previous studies with a detailed
 investigation on the energy and centrality dependence of $v_{2}$ of 
 light nuclei with more event statistics. During the BES program, the
 STAR experiment has taken data over a wide range of collision energies
 from $\sqrt{s_{NN}}$ = 7.7 GeV to 200 GeV. In this paper we present
 the measurement of $v_{2}$ at mid-rapidity $(|y|<1.0)$ for light nuclei 
\makebox[\linewidth][s]{d, t,  $^{3}$He ($\sqrt{s_{NN}}$ =  200, 62.4, 
 39, 27, 19.6, 11.5,  and} 7.7 GeV), and  anti-nuclei 
$\overline{\rm d}$ ($\sqrt{s_{NN}}$ =  200, 62.4, 39,
 27, and 19.6 GeV) and $^{3}\overline{\rm He}$ ($\sqrt{s_{NN}}$ = 200 GeV)
 
 The paper is organized as follows. Sec. II briefly describes the
 experimental setup, the detectors and the particle (and light-nuclei)
 identification (PID) techniques. The centrality definition, event
 selection, event plane reconstruction and the event plane resolution
 correction are also discussed, along with the extraction procedure of
 light-nuclei $v_{2}$. Presented in Sec. III are the $v_{2}$ results
 for minimum bias collisions, the centrality dependence, and a
 physical interpretation of the results. A comparison between
 light-nuclei $v_{2}$ measured in this experiment and those calculated
 from blast wave and transport-plus-coalescence model is also
 shown. Sec. IV summarizes the physics observations and discusses the
 main conclusions from the results.  
 
%-------------------- dE/dx plot ---------------------
 \begin{figure*}[t]
 \centering
 \includegraphics[totalheight=9.0cm]{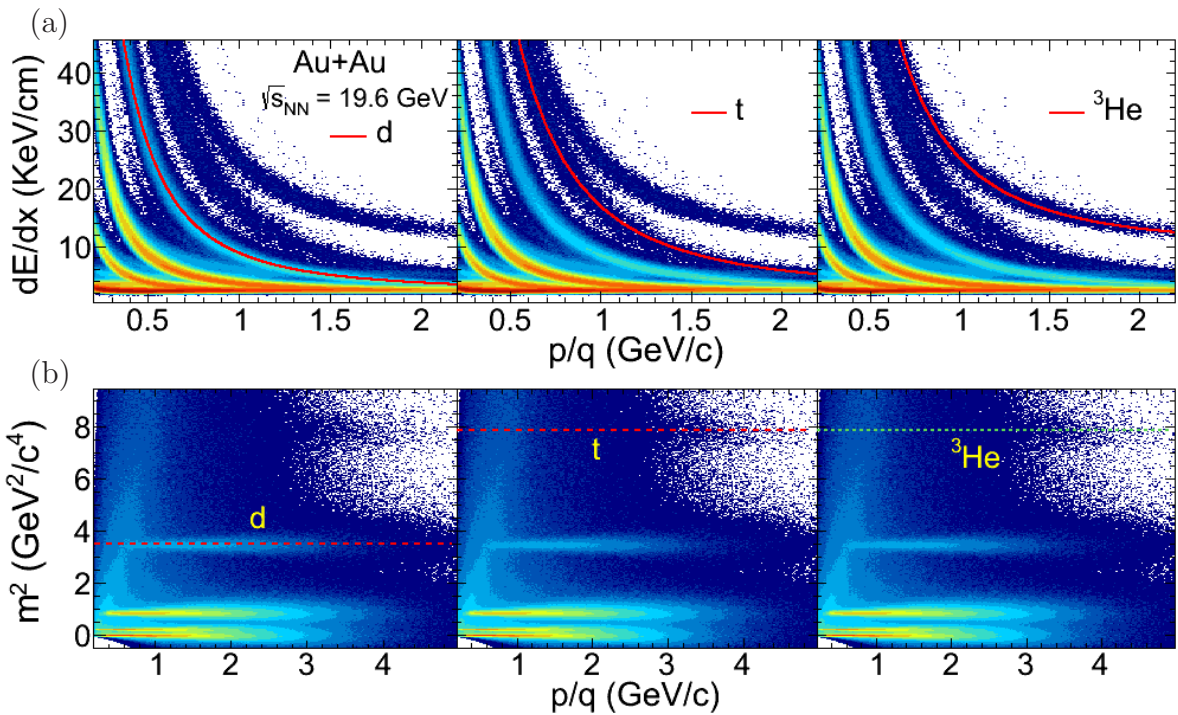}
 \put(-434,258){\large{(a)}} 
 \put(-434,125){\large{(b)}} 
 \caption{\small{(color online) (a) Specific energy loss (dE/dx) as a
     function of rigidity (momentum/charge). Theoretical dE/dx
     expectations, using the model in Ref.~\cite{Bichsel} of d, t,
     $^{3}$He are shown by solid curves. (b) Mass squared ($m^{2}$)  
     as a function of momentum for mid-rapidity charged particles.  
     The dotted lines correspond to $m^{2}$ of different nuclei. Both
     results are from minimum bias Au+Au collisions at $\sqrt{s_{NN}}$
     = 19.6 GeV. }} 
\label{fig:dedx}
\end{figure*}

 \section{Experimental setup}

  STAR is a multipurpose experiment at the RHIC facility at
 Brookhaven National Laboratory. It consists of a longitudinally-oriented
 (beam direction) solenoidal magnet and
 a collection of detectors for triggering, PID, and
 event categorization~\cite{STAR_nim}. The main detectors used for
 this analysis are the Time Projection Chamber (TPC)~\cite{STAR_TPC} 
 and the Time of Flight (TOF) detector~\cite{STAR_ToF}. The following
 subsections briefly describe their operations and PID techniques.

 \subsection{TPC Measurements} 

 The TPC is the primary tracking
 device in the STAR experiment which uses ionisation in a
 large gas volume to detect trajectories of charged particles.
 Curvature in the solenoidal field enables determination of the
 charge sign and rigidity (momentum/charge). The TPC has
 full azimuthal coverage and a uniform pseudorapidity range of
 $|\eta|<1.0$~\cite{STAR_TPC}. The TPC can record up to 45 hit
 positions and specific ionisation energy loss (dE/dx) samples along 
 tracks. Truncated means of the dE/dx samples are used
 for PID by comparing to theoretical expectations, using improved
 Bethe-Bloch functions~\cite{Bichsel}, at the measured rigidities to
 characterize the probability for being any particular species.
 PID consequently allows deduction of the particles' charges and
 momenta. A representative plot of measured track dE/dx
 versus rigidity is shown in Fig.~\ref{fig:dedx}(a)
 for minimum bias (defined later) Au+Au collisions at $\sqrt{s_{NN}}
 =$ 19.6 GeV. The
 theoretical curves are shown as solid lines. 
 Primary collision vertices are found through
 fits involving candidate daughter tracks, and a typical central
 collision at the top RHIC energy (with perhaps $\sim$1000
 reconstructed tracks) may achieve a vertex position resolution of
 $\sim$350 $\mu$m. These daughter tracks are then refitted using
 their vertex as a constraint to create a collection of primary tracks.

\subsection{TOF Measurements} 

The TOF detector \cite{STAR_ToF} in STAR uses
Multigap Resistive Plate Chambers (MRPCs) and was fully
installed in the year 2010. It covers 2$\pi$ in azimuth
within the pseudorapidity interval $|\eta|$$<$0.94. The TOF
detector and the Vertex Position Detector (VPD) \cite{STAR_VPD}
measure the time interval, $t$, over which a particle
travels from the primary collision vertex to a read-out
cell of the TOF detector. This time interval information is
combined with the total path length, $S$, measured by the
TPC to provide the inverse velocity, 1/$\beta$, via
1/$\beta$$=$c$t$/$S$, where c is the speed of light. The
track mass-squared is then given by
$m^{2}$$=$$p^{2}$(1/$\beta^{2}$$-$1).
For collision energies below $\sqrt{s_{NN}}$$=$ 39 GeV the VPD
efficiency is too low to use in every event. Instead, for these data
sets a start time for each collision is inferred by working backwards
from the TOF-measured stop times of a very limited selection of
particles which are very cleanly identified in the TPC. 
The total time interval resolution obtained of 90-110 ps
results in PID capabilities that are complementary to those from the
TPC dE/dx at low momenta and also extend to momenta of several GeV. A
representative plot of $m^{2}$ as a function of the particle momentum
is shown in Fig.~\ref{fig:dedx}(b) for minimum bias Au$+$Au collisions
at $\sqrt{s_{NN}}$$=$19.6 GeV. As the mass of a particle is a 
constant quantity, we expect horizontal bands for individual
(anti-)nuclei as shown by the dotted lines in Fig.~\ref{fig:dedx}(b). 
We have selected individual nuclei using the $m^{2}$ which lie within  
3$\sigma$ from the constant mean (dotted line).

%-------------------- resolution plot --------------
\begin{figure*}[t]
 \vspace{0.1cm}
 \centering
 \includegraphics[totalheight=8.0cm]{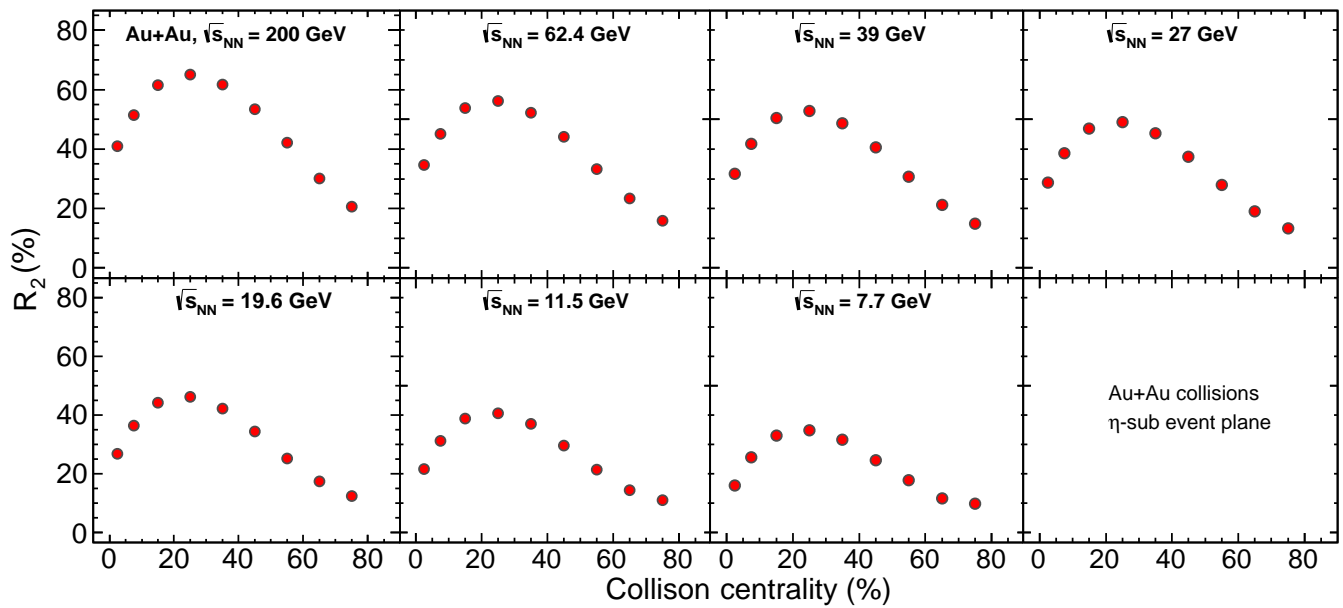}
 \caption{\small {Resolution correction factor ($R_{2}$) of sub-event planes as
  a function of centrality for Au+Au collisions at $\sqrt{s_{NN}}$  = 200,
 62.4, 39, 27, 19.6, 11.5, and 7.7 GeV.} }
\label{fig:reso}
\end{figure*}

 \subsection{Trigger and event selection} 

 The minimum bias events for all of the collision energies
 are based on a coincidence of the signals from the Zero-Degree
 Calorimeters (ZDC)~\cite{STAR_ZDC}, VPD, and/or Beam-Beam Counters
 (BBC)~\cite{STAR_BBC}. Due to larger beam emittance at lower collision
 energies, Au$+$Au-triggered events are contaminated with
 Au$+$beam-pipe events. The radius of the beam-pipe going through the
 center of the TPC is 3.95 cm. Therefore, such Au$+$beam-pipe events
 are removed by requiring the primary vertex position to be within a
 transverse radius of less than 2 cm in the
 XY-plane~\cite{v2_BES_prc}. The $z$-position of the primary vertices
 (vertex-${z}$) is limited to the values listed in
 Table~\ref{table:events}~\cite{v2_BES_prc} to ensure good quality
 events. 

%---------------- MB events table -----------------
\begin{table}[h]
 \small
 \caption{The vertex-${z}$ acceptance and total number of minimum-bias
   (MB) events for each energy ($\sqrt{s_{NN}}$).}
 \vspace{0.4cm}
\begin{tabular}{ c  c  c  c }
 \hline
 \hline  
  $\sqrt{s_{NN}}$ (GeV)  & vertex-${z}$ (cm)  & MB events ($\times10^{6}$)  \\ [0.6ex]
 \hline
 200 & $\mid$vertex-$z\mid\ <$ 30    &  241 \\ [0.2ex] 
 
 62.4 & $\mid$vertex-$z\mid\ <$ 40   & 62 \\   [0.2ex] 
 
 39   &  $\mid$vertex-$z\mid\ <$ 40   & 119 \\ [0.2ex] 

 27   &  $\mid$vertex-$z\mid\ <$ 70   & 60  \\  [0.2ex] 

 19.6 &  $\mid$vertex-$z\mid\ <$ 70   & 33 \\   [0.2ex] 

 11.5 &  $\mid$vertex-$z\mid\ <$ 50   & 11 \\   [0.2ex] 

  7.7 &  $\mid$vertex-$z\mid\ <$ 70  &   4 \\    [0.2ex] 
 \hline
\hline
 \end{tabular}
 \label{table:events}
 \end{table}

 Furthermore, an extensive quality assurance of the events    
 was performed based on the mean transverse momenta, the mean vertex
 position, the mean interaction rate, and the mean multiplicity in the
 detector. Run periods were removed if one of those quantities was
 more than 3$\sigma$ away from the global mean value. The total number
 of minimum bias events used in this analysis after these quality
 assurance cuts for each collision energy are shown in Table~\ref{table:events}.

\subsection{Centrality definition} 

 The centrality of each event is defined based on the
 uncorrected charged particle multiplicity
 ($dN_{events}/dN_{charge}^{raw}$) distribution, where $N_{events}$
 is the number of events and $N_{charge}^{raw}$ is the number of 
 charged particles measured within $|\eta| <$ 0.5~\cite{v2_BES_prc}. 
 Thus, for example, 0\%-5\% central events correspond to the events in
 the top 5\% of the multiplicity distribution. The charged particle
 multiplicity distributions for all energies can be described by a
 two-component model~\cite{two_comp}. The two-component model is a
 Glauber Monte-Carlo simulation in which the multiplicity per unit
 pseudo-rapidity ($dN_{charge}/d\eta$) depends on the two components,
 namely, number of participant nucleons ($N_{part}$) and number of
 binary collisions ($N_{coll}$): 

 \begin{equation}
 \frac{dN_{charge}}{d\eta}\ =\ n_{pp}[(1-x)\frac{N_{part}}{2}\ +\ xN_{coll}].
 \label{eq:two_comp}
 \end{equation}

\noindent  The fitting parameter $n_{pp}$ is the $dN_{charge}/d\eta$ in
 minimum-bias p$+$p collisions and $x$ is the fraction of produced
 charged particles from the hard component. The centrality class is 
 defined by calculating the fraction of the total cross-section
 obtained from the simulated multiplicity.  Due to trigger inefficiencies,
 many of the most peripheral events were not recorded. This results in
 a significant difference between the measured distribution of charged
 particle multiplicities and the Glauber Monte Carlo (MC) simulation
 for peripheral collisions. When determining $v_{2}$ in a bin of
 multiplicity wide enough to see variation in the trigger inefficiency
 across the bin (e.g. for a minimum bias measurement), it is necessary
 to compensate for this variation by weighting particle yields in each
 event by the inverse of the trigger efficiency at that event's
 multiplicity~\cite{v2_BES_prc}.  The correction is about 5\% for the
 peripheral (70\%-80\%) events, and becomes negligible for central
 events. However, the corrections are severe for 80\%-100\%
 central events. Therefore, 80\%-100\% central events are not included
 in the current analysis, and minimum bias is defined for all data
 presented here as 0\%-80\%. In addition to the trigger inefficiency,
 two additional corrections are also applied to account for the
 vertex-${z}$ dependent inefficiencies. These corrections account for the
 acceptance and detector inefficiencies and the time-dependent changes  
 in $dN_{events} /dN_{charge}^{raw}$. 

%---------------- Z-plot   &   phi-psi2 plot ------------------
 \begin{figure*}[t]
 \centering
 \includegraphics[totalheight=4.0cm]{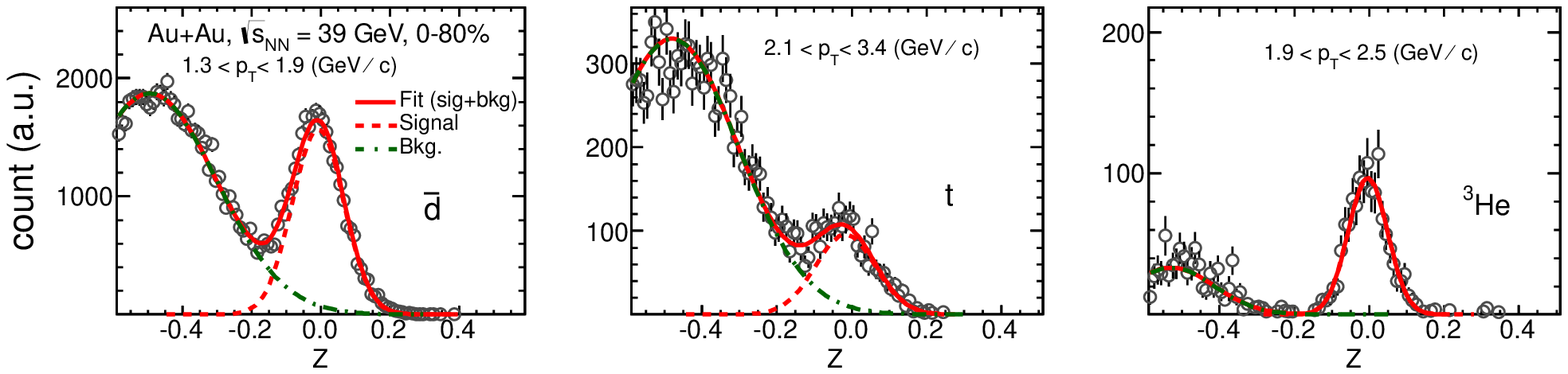}
 \put(-438,106){\large{(a)}} 
 \break
 \includegraphics[totalheight=4.0cm]{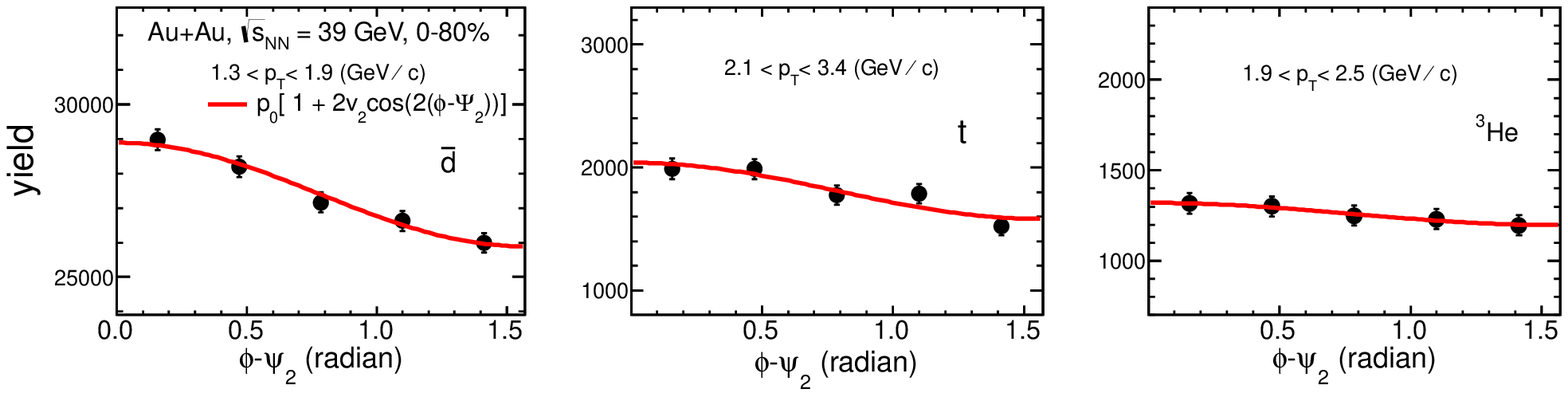}
 \put(-438,106){\large{(b)}} 
 \caption{\small{(color online) (a) $Z$ distributions for mid-rapidity
     $\overline{\rm d}$, t, $^{3}$He. The different $p_{T}$ ranges are
     for acceptance-representative purpose. The $Z$ distribution for
     each species is fitted with a two-Gaussian function. One Gaussian
     is used to describe the $Z$ distribution for the species of
     interest (dashed line), and another Gaussian is used to describe
     the background (dot-dashed line). (b) $(\phi-\Psi_{2})$
     distributions for mid-rapidity $\overline{\rm d}$, t, and
     $^{3}$He. Solid lines are fitted 2$^{nd}$ order Fourier
     functions. All plots use minimum bias Au+Au collisions at
     $\sqrt{s_{NN}}$ = 39 GeV.}} 
 \label{fig:Zdist}
 \end{figure*}

 \subsection{Event plane and resolution correction} 

 The azimuthal distribution of produced particles with
 respect to reaction plane angle $(\Psi_{r})$ can be expressed  in
 terms of a Fourier series,

 \begin{equation}
 \frac{dN}{d(\phi-\Psi_{r})} \propto 1+ 2v_{1}\cos(\phi-\Psi_{r}) + 2v_{2}\cos(2(\phi-\Psi_{r})) + ... 
 \label{eq:flow}
 \end{equation}

 \noindent where $\phi$ is the azimuthal angle of the produced particle.
 $\Psi_{r}$ is defined as the angle between $x$-axis in the laboratory
 frame and axis of the impact parameter. Because we cannot directly
 measure $\Psi_{r}$, we must use a proxy.
 The second order azimuthal anisotropy or elliptic flow $(v_{2})$ is
 measured with respect to the $\rm{2}^{nd}$ order event plane angle
 ($\Psi_{2}$) instead.  $\Psi_{2}$ is calculated using the azimuthal
 distribution of all   reconstructed primary tracks ($N$)~\cite{art1}:

  \begin{equation}
  \Psi_{2}\ = \frac{1}{2}\tan^{-1}(\frac{Q_{2,y}}{Q_{2,x}}) .
  \label{eq:psi2}
 \end{equation}

 $Q_{2x}$ and $Q_{2y}$ are defined as

 \begin{subequations}
 \label{sub:QxQy}
  \begin{align}
   Q_{2}\cos(2\Psi_{2})\ =\ Q_{2,x} = \sum\limits_{i=1}^{N} w_{i}\cos(2\phi_{i}) , \\
   Q_{2}\sin(2\Psi_{2})\ =\ Q_{2,y} = \sum\limits_{i=1}^{N} w_{i}\sin(2\phi_{i}) ,
  \end{align}
 \end{subequations}

 \noindent where $w_{i}$ are the weights which
 optimise the event plane resolution~\cite{art1}. In
 this analysis, the weights scale with track-$p_{T}$, then
 saturate above 2.0 GeV/$c$.
 To reduce biases due to short range correlation, we
 utilize the sub-event plane method~\cite{art1}. In this
 analysis, the two sub-events were defined in $\eta$ windows 
 of $\eta^{-}$ (-1.0~$<$~$\eta$~$<$~-0.05) and $\eta^{+}$ 
 (0.05~$<$~$\eta$~$<$~1.0). Event plane angles are calculated within
 each $\eta$ window, $\Psi_{2\eta^{-}}$ and 
 $\Psi_{2\eta^{+}}$ respectively, and $v_{2}$ is calculated
 in each sub-event using the opposite sub-event's event
 plane angle. The $\eta$ gap ($\Delta
 \eta$ = 0.1) between the sub-events reduces the short range non-flow 
 contributions and avoids the self-correlation. 
 %to the $v_{2}$ measurements.  
 However, long range correlations may persist~\cite{art2}.   

       Due to the acceptance inefficiency of the detectors, the
 reconstructed event plane distributions are not uniform.
 Therefore, we apply event-by-event recenter~\cite{recenter}  
 and shift~\cite{shift} corrections. Finite multiplicities also
 restrict the degree to which the found event plane angles coincide
 with the true reaction plane angle. Hence, a
 resolution correction is applied to the observed elliptic flow
 ($v_{2}^{obs}$): $v_{2}$ = $v_{2}^{obs}/R_{2}$. 
 We determine the resolution correction factor ($R_{2}$)
 in the $\eta$ sub-event plane method as follows~\cite{art1}:

 \begin{equation}
   R_{2}\ =\ \sqrt{<\cos[2(\Psi_{2\eta{+}}-\Psi_{2\eta^{-}})]>} .
 \label{eq:reso_etasub}
 \end{equation}

\noindent The resolution as a function of centrality for $\eta$ sub-event
 planes is shown in Fig.~\ref{fig:reso} for Au+Au collisions. $R_{2}$
 grows with increasing multiplicity (which is small for peripheral collisions)
 and with increasing $v_{2}$ (which is small for the most central collisions),
 so its value peaks in mid-central (20\%-30\%) collisions where neither is small.

%-------------------- v2 minimum bias --------------------------
 \begin{figure*}[t]
 \centering
 \includegraphics[totalheight=8.0cm]{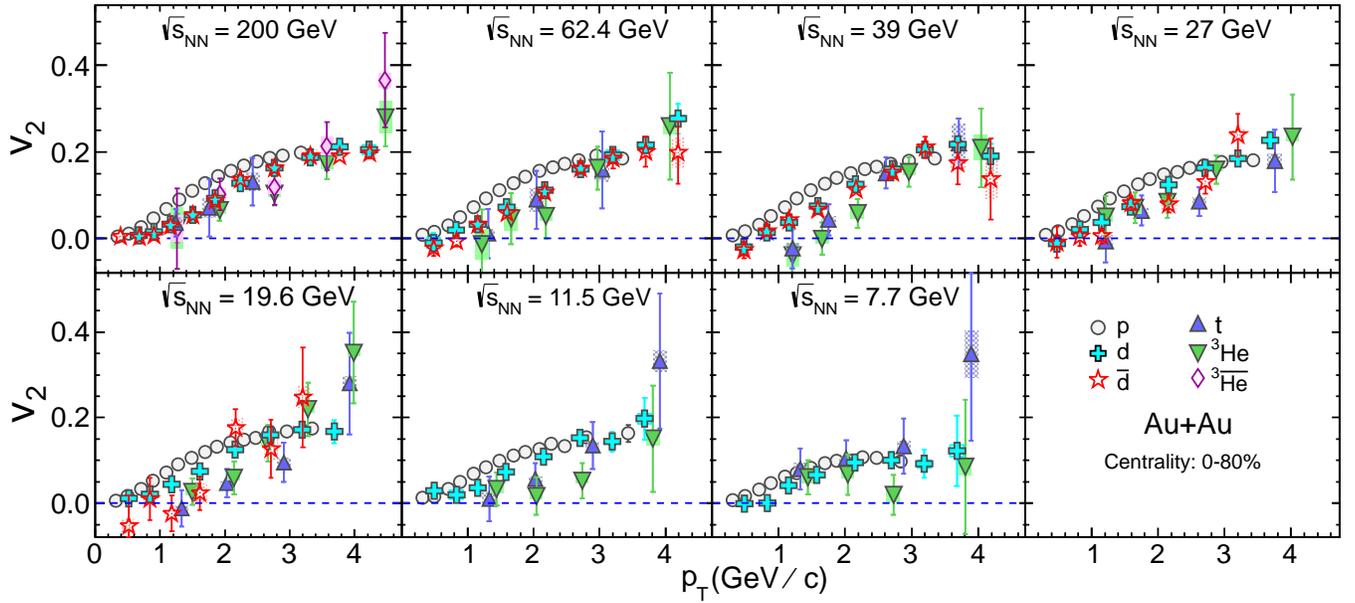}
 \caption{\small {(color online) Mid-rapidity $v_{2}(p_{T})$ for
 d, $\overline{\rm d},$ t, $^{3}$He, and $^{3}\overline{\rm He}$ from minimum 
 bias Au+Au collisions at $\sqrt{s_{NN}}$ = 200, 62.4, 39, 27, 
 19.6, 11.5, and 7.7 GeV. For comparison, proton $v_{2}(p_{T})$ are also
 shown as open circles~\cite{v2_BES_prc,v2_200_QM}. Lines
 and boxes at each marker represent statistical and systematic errors respectively.}}
 \label{fig:v2_pt}
 \end{figure*}

\subsection{Extraction of yield and $v_{2}$ of nuclei} 
%--------------------d-dbar v2 difference------------------
To identify light nuclei, we define a variable $Z$ such that 

\begin{equation}
 Z\ =\ \ln[\rm{(dE/dx)}_{expt} / \rm{(dE/dx)}_{theory}],
\end{equation}

\noindent where $\rm{(dE/dx)}_{expt}$ is the energy loss of the 
light nuclei measured by the TPC detector in the experiment and
 $\rm{(dE/dx)}_{theory}$ is the theoretical energy loss as obtained
 from the modified Bethe-Bloch formula~\cite{Bichsel}.
 After cutting on $m^2$ from TOF (see Fig.~\ref{fig:dedx}(b)) to
 reduce backgrounds under the signals, the yields are extracted
 from the $Z$ distributions in various $p_{T}$ and $(\phi-\Psi_{2})$
 bins for each species of interest with a two-Gaussian function (one
 for the signal, the other for the
 background). Figure~\ref{fig:Zdist}(a) 
\makebox[\linewidth][s]{shows sample $Z$ distributions 
 for $\overline{\rm d},\ t$ and  $^{3}$He,}
\makebox[\linewidth][s]{respectively, within  0
  $<(\phi-\Psi_{2})<\pi/10$ for} 
\makebox[\linewidth][s]{1.3 $<p_{T}<$ 1.9 GeV/$c$, 2.1 $<p_{T}<$ 3.4
  GeV/$c$, and} 
 1.9  $<p_{T}<$ 2.5 GeV/$c$ for minimum bias Au+Au data at
 $\sqrt{s_{NN}} = 39$ GeV.  
 The azimuthal angle variation of this yield is then fitted with a
 2$^{nd}$ order Fourier function to get the elliptic flow coefficient
 ($v_{2}^{obs}$). Figure~\ref{fig:Zdist}(b) shows the $(\phi-\Psi_{2})$
 distributions for $\overline{\rm d}$, t and $^{3}$He for the same
 $p_{T}$ ranges as shown for $Z$ distributions in
 Fig.~\ref{fig:Zdist}(a). As the ($\phi-\Psi_{2}$) distribution is
 expected to be symmetric about 0 and $\pi/2$, the data points have
 been folded onto 0-$\pi/2$  to reduce the statistical errors. 

 The fitted 2$^{nd}$ order Fourier functions are shown 
 in Fig.~\ref{fig:Zdist}(b). Event plane resolution
 correction factors are determined in each centrality bin.
 For $v_{2}$ integrated over multiple centrality bins,
 species-yield-weighted mean of the individual centrality bins' resolutions
 are used: $v_{2}\ =\ v_{2}^{obs} \langle\frac{1}{R_{2}}\rangle$~\cite{hiroshi_alex}.

\subsection{Calculation of systematic uncertainty and removal of
  beam-pipe contaminations}
We have reduced light-nuclei contaminants from interactions with the
beam pipe by cutting tightly on the projected distance of closest
approach (DCA) to the primary vertex. Remaining contaminants from such
interactions are removed statistically by fitting the DCA distribution
of nuclei with that of anti-nuclei (which are expected to have no such
background) in each ($\phi - \Psi_{2}$) bin.
Systematic uncertainties are determined by varying cuts
used in particle identification and background rejection, and by
varying fitting methods and ranges when measuring yields. 
The absolute magnitude of uncertainties range from 2\%-5\% for
intermediate $p_T$ (1.0 $< p_{T} <$ 3.0 GeV/c) and from 5\%-8\% 
for low and high $p_T$.
 
%-------------------- v2 mass ordering--------------------------
 \begin{figure*}[t]
\centering
 \includegraphics[totalheight=8.0cm]{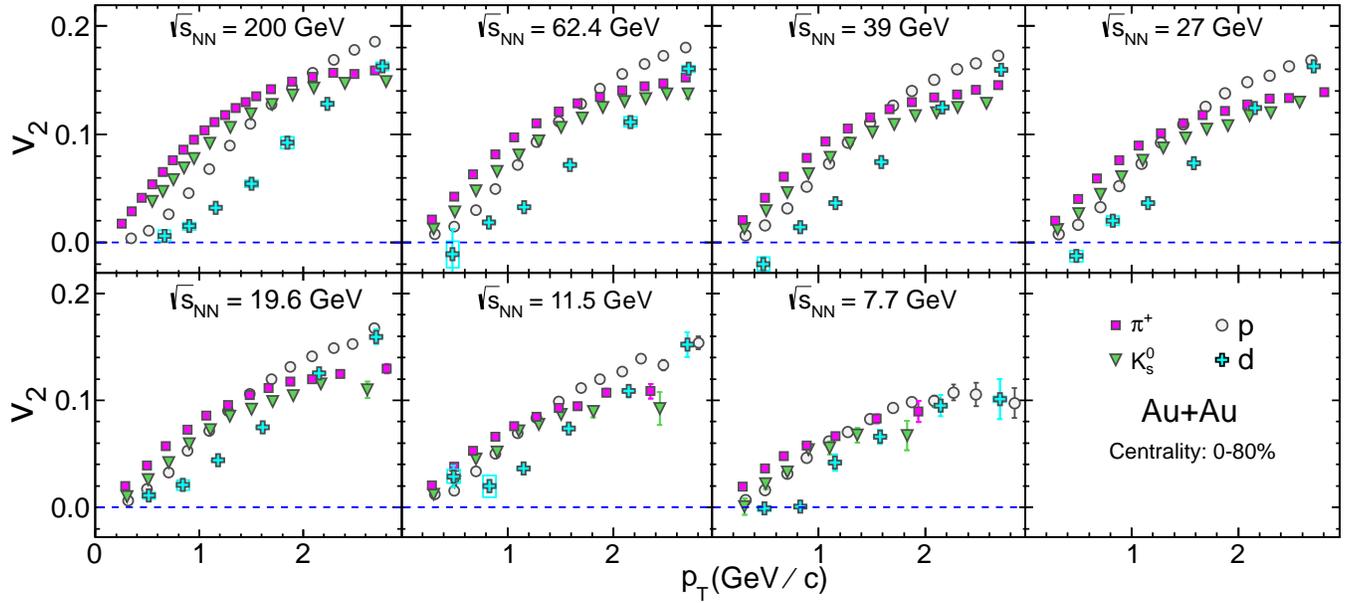}
 \caption{\small {(color online) Mid-rapidity $v_{2}(p_{T})$ 
     for $\pi^{+}$ (squares), $\rm K_{s}^{0}$ (triangles), p (open circles), and d
     (crosses) for minimum bias Au+Au collisions at
     $\sqrt{s_{NN}}$ = 200, 62.4, 39, 27, 19.6, 11.5, and 7.7 GeV.} }
 \label{fig:mass_ordering}
\end{figure*}
 
 \begin{figure*}[t]
 \vspace{0.4cm} 
 \includegraphics[totalheight=8.0cm]{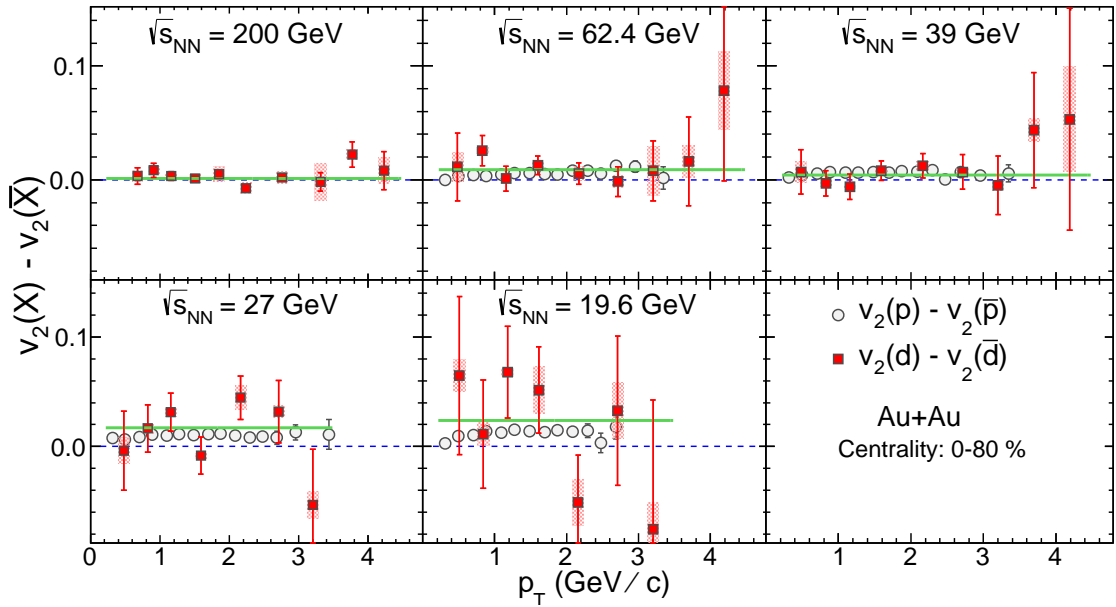}
 \caption{\small{(color online) The difference in $v_{2}$ of d and 
     $\overline{\rm d}$ as a function of $p_{T}$ for minimum bias Au+Au
     collisions at $\sqrt{s_{NN}}$ = 200, 62.4, 39, 27, and 19.6 GeV,
     along with differences between p and $\overline{\rm p}$~\cite{v2_BES_prc}.
     Solid lines correspond to constants fit to the data (see text for details).}} 
\label{fig:v2_ddbar}
\end{figure*}

%-----------  centrality dependence -------------
 \begin{figure*}[t]
 \centering
 \includegraphics[totalheight=7.5cm]{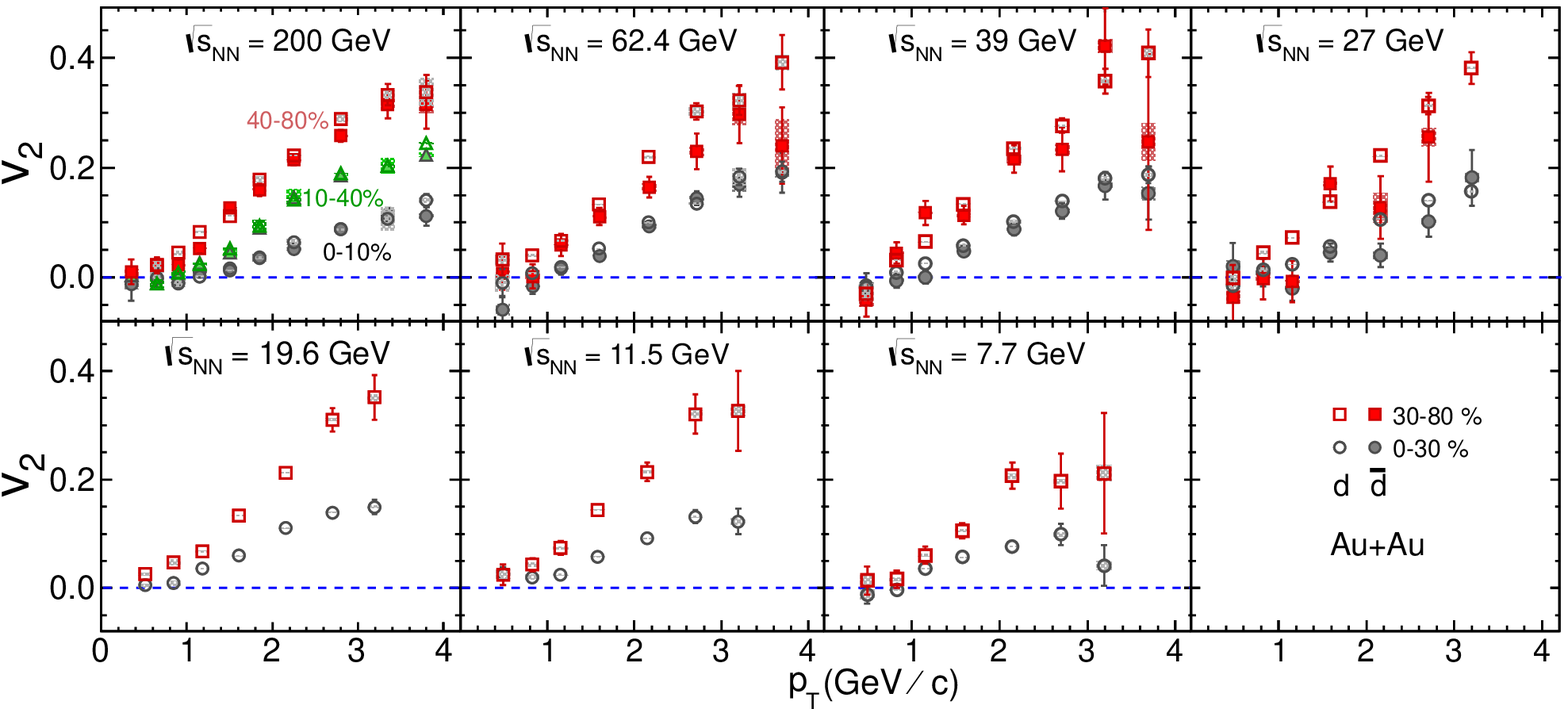}
 \caption{\small{(color online) Centrality dependence of mid-rapidity
     $v_{2}(p_{T})$ of d (open markers) for Au+Au collisions at
     $\sqrt{s_{NN}}$ = 7.7 - 200 GeV and $\overline{\rm d}$ (solid
     markers) for $\sqrt{s_{NN}}$ = 27 - 200 GeV. For $\sqrt{s_{NN}}$
     = 200 GeV,  circles correspond to 0\%-10\%, triangles to
     10\%-40\%, and squares to 40\%-80\% central events.
     For other collision energies, circles correspond to
     0\%-30\% and squares to 30\%-80\% central events.}}
\label{fig:v2_cent}
\end{figure*}

\section{Results and discussion}

 \subsection{General properties of $v_{2}(p_{T})$} 

 Figure~\ref{fig:v2_pt} shows the energy dependence of the 
 $v_{2}$ of the light (anti-)nuclei d, $\overline{\rm d}$, t, $^{3}$He,
 and $^{3}\overline{\rm He}$ as a function of $p_{T}$ for minimum bias
 Au+Au collisions. Insufficient statistics preclude measuring
 differential anti-nuclei $v_{2}$ at several collision energies. 
 The $v_{2}(p_{T})$ of all light-nuclei species and anti-nuclei
\makebox[\linewidth][s]{species ($\overline{\rm d}$ at $\sqrt{s_{NN}}$ = 19.6 $-$ 200 
 GeV and}
 $^{3}\overline{\rm He}$ at $\sqrt{s_{NN}}$ = 200 GeV) show
 monotonically increasing trend with increasing $p_{T}$
 (Fig.~\ref{fig:v2_pt}). Mass ordering of $v_{2}(p_{T})$ for $p_{T} <
 $ 2.0 GeV$/c$ is clear in both Figs.~\ref{fig:v2_pt}
 and~\ref{fig:mass_ordering}, where the $v_{2}(p_{T})$ of $\pi^{+}$,
 K$^{0}_{s}$, and p from Ref.~\cite{v2_BES_prc,v2_200_QM} are also
 included (heavier species have a lower $v_{2}$ in this $p_{T}$
 range). Such ordering occurs naturally in a hydrodynamic $+$
 coalescence model of heavy-ion collisions~\cite{hydro}. The negative
 $v_{2}$ observed for some (anti)-nuclei could be the result of radial
 flow.
 
 Figure~\ref{fig:v2_ddbar} presents the difference of $v_{2}(p_{T})$
 between d and $\overline{\rm d}$ ($\Delta v_{2}$), along with the
 difference between p and $\overline{\rm p}$ for
 comparison~\cite{v2_BES_prc,v2_200_QM}. Statistical uncertainties are 
 too large to draw conclusions about any collision energy dependence,
 but the $\Delta v_{2}$ data are qualitatively consistent with the
 (anti-)protons and the results of fitting a constant at each energy
 (solid lines in Fig.~\ref{fig:v2_ddbar}) are consistently positive:
 %and hint at a possible rise at the lower energies: 
 0.0012$\pm$0.0014, 0.009$\pm$0.005, 0.0044$\pm$0.0046,
 0.017$\pm$0.009, 0.024$\pm$0.019 for $\sqrt{s_{NN}}$ = 200, 62.4,
 39, 27, and 19.6 GeV respectively. 

 Figure~\ref{fig:v2_cent} shows $v_{2}(p_{T})$ of d and
 $\overline{\rm d}$ in 0\%-30\% and 30\%-80\% central events for
 where they could be measured in Au+Au collisions at $\sqrt{s_{NN}}$ =
 62.4 to 7.7 GeV. For 200 GeV $v_{2}$ are measured in three centralities:
 0\%-10\%, 10\%-40\%, and 40\%-80\%.
The observed centrality dependences are qualitatively similar to those
seen in identified hadrons~\cite{v2_BES_prc,v2_PID_cent}, with d and
$\overline{\rm d}$ showing similar behavior for all centralities measured.

 \subsection{Blast wave model}  
\begin{figure*}[t]
 \vspace{0.0cm}
 \centering
 \includegraphics[totalheight=9.0cm]{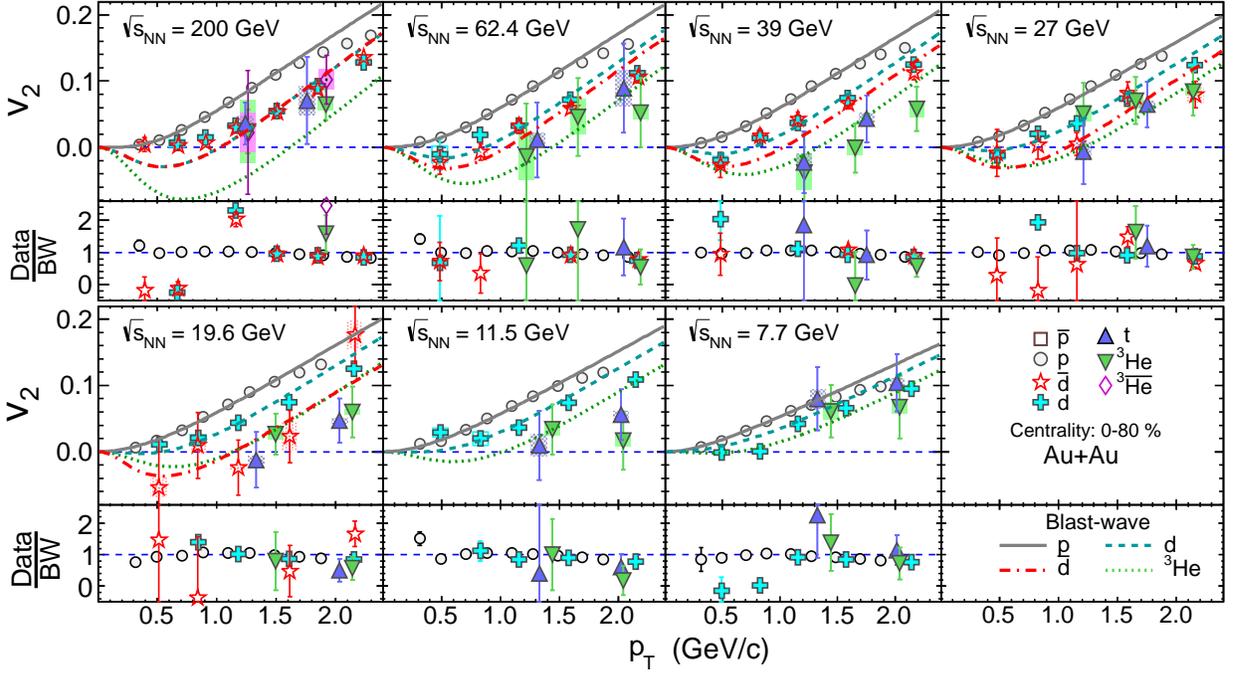}
 \caption{\small{(color online) Blast wave model predictions (lines)
     of $v_{2}$ for d, $\overline{\rm d}$, t, $^{3}$He ($^{3}\overline{\rm He}$)
     compared with the data for minimum bias  Au+Au collisions. The
     blast wave model and parameter values have been used
     from~\cite{PID_BW_fit}. (Some data points in the lower panels are
     off-scale.)}}  
 \label{fig:v2_BW}
 \end{figure*}

 The nuclear fireball model was first introduced by Westfall {\it et al.} 
 to explain midrapidity proton-inclusive spectra~\cite{Gary_fireball}.
 Later, Siemens and Rasmussen~\cite{Siemens_BW} generalized a
 non-relativistic formula by Bondorf, Garpman, and
 Zimanyi~\cite{Bondorf_BW} to explain nucleons and pions as they are
 produced in a blast wave of an exploding fireball. The blast wave model
 has evolved since then, with more parameters to describe both 
 $p_{T}$ spectra and anisotropic flow of produced 
 particles~\cite{Schnedermann_BW,Huovinen_BW,STAR_BW}.
 The blast wave parametrization modeled by the STAR
 Collaboration~\cite{STAR_BW} has been recently used to fit the $v_{2}$
 of identified particles~\cite{PID_BW_fit}. This version of blast wave
 has four parameters, namely kinetic freeze-out temperature ($T$),
 transverse expansion rapidity ($\rho_{0}$), amplitude of its azimuthal
 variation ($\rho_{a}$), and the variation in the azimuthal density of
 the source elements ($s_{2}$)~\cite{PID_BW_fit}. The fit parameters
 obtained from blast wave fits to the $v_{2}$ of identified particles
 are listed in Table 1 of~\cite{PID_BW_fit}. We have used the same
 blast wave model and fit parameter values to check whether the
 bast wave model also reproduces the $v_{2}$ of light nuclei measured
 in the data. Figure~\ref{fig:v2_BW} shows the blast wave model
 predictions for light nuclei, along with the measurements.
 As is evident from Fig.~\ref{fig:v2_BW}, blast wave
 model under-predicts the $v_{2}$ of d and $\overline{\rm d}$ at low
 $p_{T}$ ($p_{T} <$ 1.0 GeV/$c$) for most of the collision energies.
 Similar conclusion for t, $^{3}$He ($^{3}\overline{\rm
 He}$) is difficult to make due to their large statistical
 uncertainty. However, for $\sqrt{s_{NN}}$ = 200 GeV, the blast wave
 model clearly fails to reproduce the measured $v_{2}$ of light nuclei
 of all species at low $p_{T}$  ($p_{T} <$ 1.0 GeV/$c$). 

%--------------- blast wave model  ----------------

%--------------- atomic mass scaling ----------------
 \begin{figure*}[t!]
 \centering
 \includegraphics[totalheight=9.0cm]{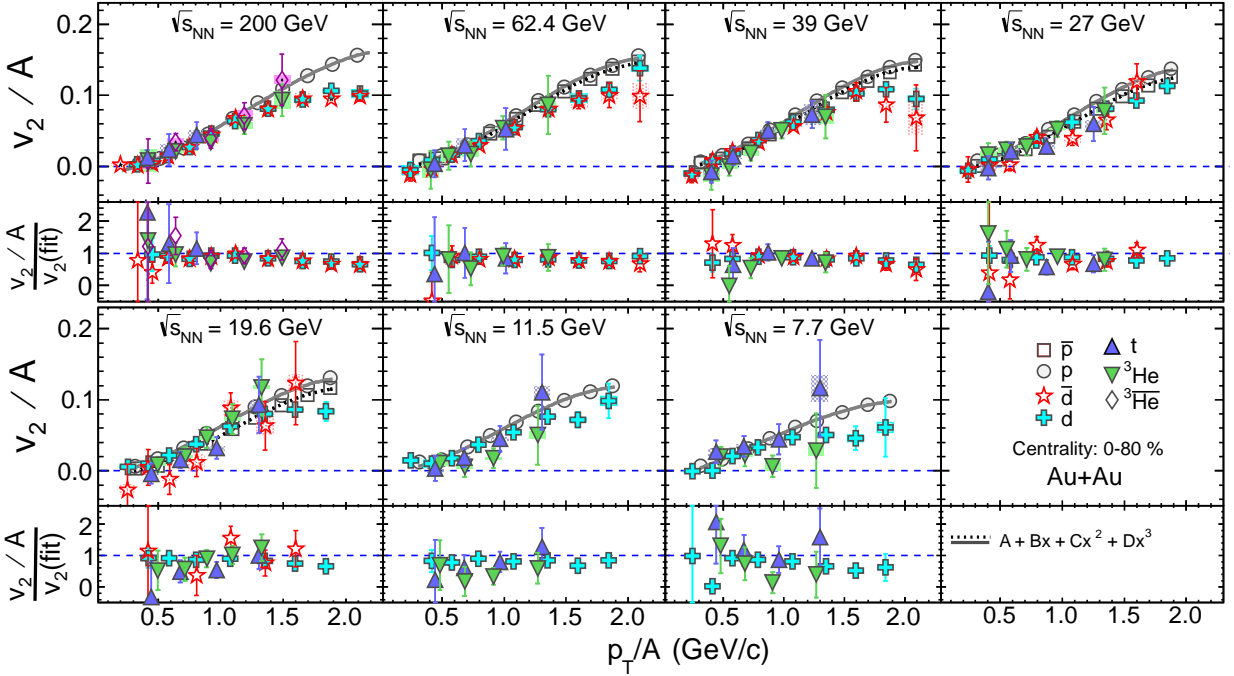}
 \caption{\small{(color online) Atomic mass number ($A$) scaling of the mid-rapidity
     $v_{2}$ of p, $\overline{\rm p}$, d, $\overline{\rm d}$,
     t, $^{3}$He, and $^{3}\overline{\rm He}$ from minimum bias 
     Au+Au collisions at $\sqrt{s_{NN}}$ = 200, 62.4, 39, 27, 19.6, 11.5,
     and 7.7 GeV. Grey Solid (black dotted) lines correspond to
     $3^{rd}$ order polynomial fits to the p ($\overline{\rm p}$)
     $v_{2}$ data. The ratios of $[v_{2}/A]/fit$ for d, $\overline{\rm
     d}$, t, and $^{3}$He are shown in the lower panels at each
    corresponding collision energy. (Some data points in the 
    lower panels are off-scale.)}} 
 \label{fig:v2_scaling}
 \end{figure*}

%left for dinner

%--------------- coalescence result ----------------
\begin{figure*}[t]
 \centering
\vspace{0.0cm}
 \includegraphics[totalheight=7.4cm]{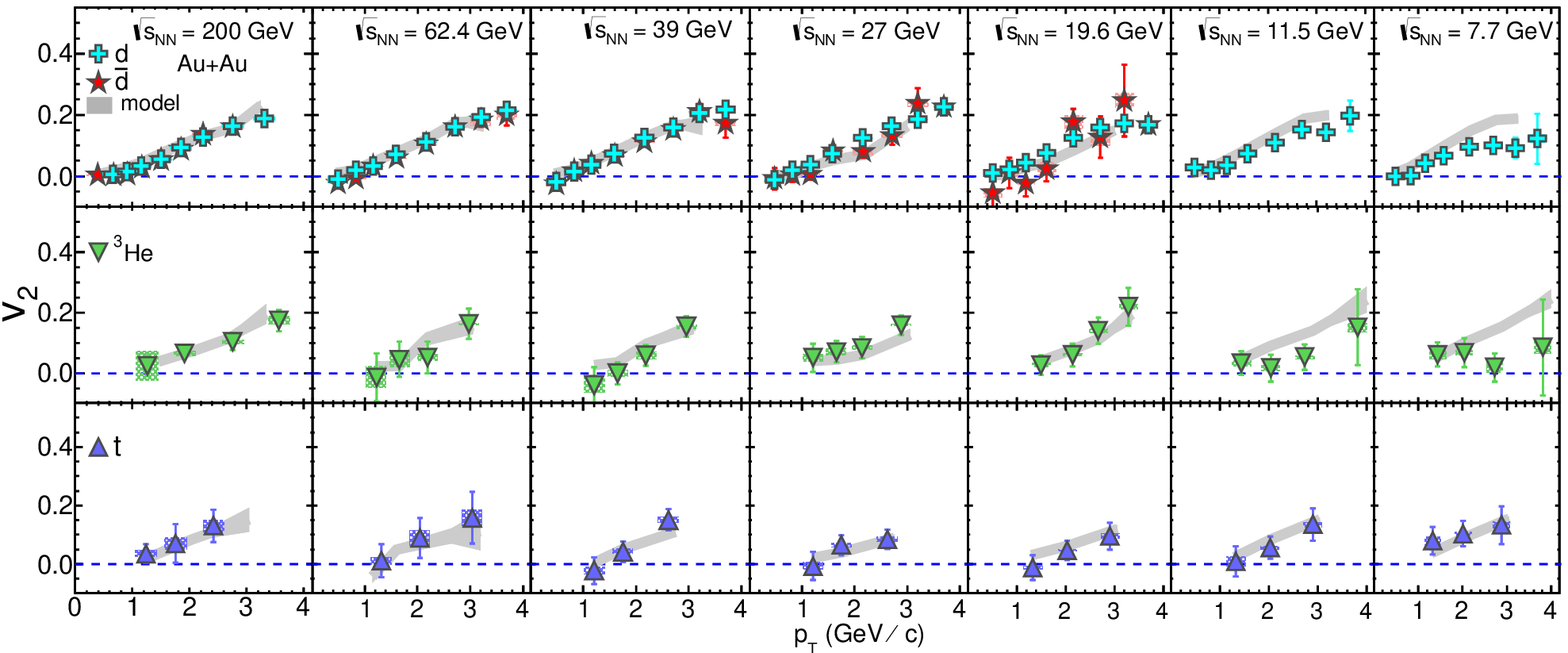}
 \caption{\small{(color online) Mid-rapidity $v_{2}$ of d, t, and $^{3}$He are
     compared with the results of AMPT+coalescence calculations
     (solid bands).} }
 \label{fig:v2_coal}
 \end{figure*}

 \subsection{Atomic mass number scaling and coalescence model}  
 Figure~\ref{fig:v2_scaling} presents the light-nuclei
 $v_{2}/A$ as a function of $p_{T}/A$, where $A$ is the atomic mass 
 number of the corresponding light nuclei. The main goal of this study  
 is to understand whether light (anti-)nuclei production
 is consistent with coalescence of (anti-)nucleons. The model
 predicts that if a composite particle is produced by coalescence
 of $n$ number of particles that are very close to each other in
 phase-space, then $v_{2}(p_{T})$ of the composite will be
 $n$ times that of the constituents~\cite{YGMA}. In
 Fig.~\ref{fig:v2_scaling} it is observed that the (anti-)nuclei
 $v_{2}/A$ closely follows $v_{2}$ of p ($\overline{\rm p}$) for
 $p_{T}/A$ up to 1.5 GeV/$c$.  The scaling behavior of these nuclei
 suggest that d ($\overline{\rm d}$) within $p_{T} < $ 3.0 GeV/$c$ and
 t, $^{3}$He ($^{3}\overline{\rm He}$) within $p_{T} <$ 4.5 GeV/$c$
 might have formed via the coalescence of nucleons
 (anti-nucleons). The low relative production of light nuclei seems to 
 favor the coalescence formalism rather than other methods, such
 as thermal production which can reproduce the
 measured particle ratios in data~\cite{CFO_nucl}. As protons and
 neutrons have the same $v_{2}$, expected from NCQ scaling, then we
 can readily see that the $v_{2}$ of t and $^{3}$He will be the same
 as they have the  same atomic mass number ($A$ = 3). We find that,
 within statistical errors, our measurement of $v_{2}$($p_{T}$) for t
 and $^{3}$He confirms this assumption. Although simple $A$ scaling
 seems to hold for the collision energies presented, the actual mechanism
 might be a more dynamic process including production and
 coalescence of nucleons in the local rest frame of the fluid
 cell. This scenario might give rise to deviations from simple $A$
 scaling.  

 It is arguable that light nuclei could have also formed via
 coalescence of  quarks as the scaling behavior holds when $v_{2}$ and
 $p_{T}$ are scaled by number of constituent quarks (e.g. 6 for d,
 $\overline{\rm d}$ and 9 for t, $^{3}$He) instead of mass number. 
 Although this process seems physically acceptable, the survival of
 light nuclei, with their low binding energies ($\sim$few MeV), is
 highly unlikely under the high temperatures requisite for
 dissociating nucleons into quarks and gluons. 
 
 To further verify the applicability of nucleon coalescence into 
 light nuclei in heavy-ion collisions, we have run the string-melting version
 of A Multi Phase Transport  (AMPT, version v1.25t7d)~\cite{AMPT}
 model of the collisions in conjunction with a dynamic coalescence
 model. The AMPT model has been used to reproduce charged particle
 multiplicity, transverse momentum spectra at RHIC and LHC, as well as 
 $v_{2}$ of identified particles at RHIC~\cite{AMPT}.
 The dynamic coalescence model has been used extensively at both
 intermediate~\cite{coal_inter} and high energies~\cite{coal_high}. 
 In this model, the probability for producing a cluster is determined
 by the overlap of the cluster's Wigner phase-space density with the
 nucleon phase-space distribution at freeze-out procured
 from AMPT. For light nuclei, the Wigner phase-space densities are
 obtained from their internal wave functions, which are taken to be
 those of a spherical harmonic oscillator~\cite{spherical}. 
 For the coalescence model we have used radii of 1.96, 1.61, and 1.74
 fm for d, t, and $^{3}$He respectively~\cite{rms_radii}. These
 parameters are kept fixed for the collision energy range presented. 
 The model's results for $v_{2}$ of d, t, and $^{3}$He  are shown as
 solid bands in Fig.~\ref{fig:v2_coal}. The data and model agree within
 errors over nearly all energies and $p_{T}$ measured, supporting 
 the theory that light nuclei are produced via nucleon coalescence in 
 heavy-ion collisions. Recently the ALICE
 collaboration has measured production of d and $\overline{\rm d}$ in
 Pb+Pb collisions at $\sqrt{s_{NN}}$ = 2.76 TeV~\cite{ALICE_ddbar}. 
 In that study, light-nuclei spectra were found to exhibit a significant
 hardening with increasing centrality. The stiffening of light-nuclei
 spectra at ALICE could be the result of increased hard scattering,
 modified fragmentation, or increased radial flow. However, the
 analysis lacks conclusive evidence regarding the 
 production mechanism of light nuclei in heavy-ion collisions. 
 In the collision energy range presented in this paper, 
 light-nuclei production favors the coalescence model.

 \section{Summary} 
 Measurements of the 2$^{nd}$ order azimuthal anisotropy,
 $v_{2}(p_{T})$ at mid-rapidity ($|y|<1.0$) have been presented for
 light nuclei d, t, $^{3}$He (for $\sqrt{s_{NN}}$ = 200, 62.4, 39, 27,
 19.6, 11.5, and 7.7 GeV), and anti-nuclei $\overline{\rm d}$
 ($\sqrt{s_{NN}}$ = 19.6$-$200 GeV) and $^{3}\overline{\rm He}$
 ($\sqrt{s_{NN}}$ = 200 GeV). Similar to hadrons over the measured
 $p_{T}$ range, light (anti-)nuclei $v_{2}(p_{T})$ show a
 monotonic rise with increasing $p_{T}$, mass ordering at low $p_{T}$,
 and a reduction for more central collisions. It is observed that
 $v_{2}$ of nuclei and anti-nuclei are of similar magnitude for
 $\sqrt{s_{NN}}$ = 39 GeV and above. 
 The difference $\Delta v_{2}$ between d and $\overline{\rm d}$ is found to
 follow the difference between p and $\overline{\rm p}$ as a
 function of collision energy. The blast wave model is found to
 under-predict the light-nuclei $v_{2}$ measured in data.
 $^{3}$He and t nuclei show similar $v_{2}$ for all collision
 energies, and in fact all the light-nuclei $v_{2}$ generally follow
 an atomic mass number scaling, which indicates that the
 coalescence of nucleons might be the underlying mechanism of 
 light-nuclei formation in high energy heavy-ion collisions. This
 observation is further corroborated by carrying out a model-based
 study of nuclei $v_{2}$ using a transport-plus-coalescence model,
 which reproduces well the light-nuclei $v_{2}$ measured in the
 data.
  
 \section*{Acknowledgements}
We thank the RHIC Operations Group and RCF at BNL, the NERSC Center at
LBNL, the KISTI Center in Korea, and the Open Science Grid consortium
for providing resources and support. This work was supported in part
by the Office of Nuclear Physics within the U.S. DOE Office of
Science, the U.S. NSF, the Ministry of Education and Science of the
Russian Federation, NSFC, CAS, MoST and MoE of China, the National
Research Foundation of Korea, NCKU (Taiwan), GA and MSMT of the Czech
Republic, FIAS of Germany, DAE, DST, and UGC of India, the National
Science Centre of Poland, National Research Foundation, the Ministry
of Science, Education and Sports of the Republic of Croatia, and
RosAtom of Russia.  
This work is supported by the DAE-BRNS project
Grant No. 2010/21/15-BRNS/2026.

\end{document}